%% file: template.tex
\documentclass[a4paper]{article}

\usepackage{INTERSPEECH2020}

\title{Adversarially Trained Multi-Singer Sequence-To-Sequence Singing Synthesizer}
\name{Jie Wu, Jian Luan}
\address{Xiaoice, Software Technology Center Asia, Microsoft\
\email{\{jiewu, jianluan\}@microsoft.com}}

\begin{document}

\maketitle
\begin{abstract}
This paper presents a high quality singing synthesizer that is able to model a voice with limited available recordings. %adversarially trained multi-singer sequence-to-sequence singing synthesizer that is able to produce high quality singing voice with limited training data. %Moreover, the singing synthesizer has a strong expressiveness in the high-pitched voice pronunciation.
Based on the sequence-to-sequence singing model, we design a multi-singer framework to leverage all the existing singing data of different singers. To attenuate the issue of musical score unbalance among singers, we incorporate an adversarial task of singer classification to make encoder output less singer dependent.  %we incorporate an adversarial loss to encourage the encoder to encode musical score in a singer-independent manner by introducing a singer classifier based on the musical score encoding. 
Furthermore, we apply multiple random window discriminators (MRWDs) on the generated acoustic features to make the network be a GAN. %a multiple random window discriminator is incorporated as a generative adversarial net (GAN) for a realistic generation of the natural singing voice.
Both objective and subjective evaluations indicate that the proposed synthesizer can generate higher quality singing voice than baseline (4.12 vs 3.53 in MOS). Especially, the articulation of high-pitched vowels is significantly enhanced. 
\end{abstract}
\noindent\textbf{Index Terms}: singing voice synthesis, multi-singer, adversarial training, GAN, sequence-to-sequence modeling

\section{Introduction}
Singing voice synthesis (SVS) is a task to synthesize specific singer's singing voice from musical score (e.g. lyric, melody and rhythm). Recent years, many deep learning based methods have been introduced into SVS to generate high quality singing voices, such DNN~\cite{nishimura2016singing} and LSTM~\cite{kim2018korean}. In addition, auto-regressive models, like Tacotron2~\cite{shen2018natural} have been successfully applied to SVS task~\cite{blaauw2017neural, yi2019singing, valle2019mellotron}.

Although auto-regressive model can achieve high quality, it suffers from exposure bias and time-consuming inference due to the forward dependency. To avoid these issues, Blaauw~\cite{blaauw2019sequence} proposed a sequence-to-sequence (Seq2Seq) singing synthesizer based on feed-forward transformer architecture, which can generate acoustic features in parallel. Feed-forward transformer has also shown its superior performance in text-to-speech (TTS) task~\cite{ren2019fastspeech}. However, to achieve a high performance, Seq2Seq singing synthesizer requires a large amount of training data from one singer. It is hard and expensive to collect them in customization application scenario. In order to reduce the amount of training data for target singer, we expect to construct a multi-singer Seq2Seq model by leveraging many existing singing data of other singers.

To build a multi-singer singing model with limited training data, one challenge is data unbalance issue. It refers to the unbalance distribution of training data among singers, such as lyric and melody. The deviation of training data distribution among singers might be considered as the singer's identification during training. To attenuate this issue, an adversarial loss~\cite{ganin2016domain} is incorporated by employing a singer classifier to encourage the encoder to learn singer-independent representation from musical scores. Adversarial training has demonstrated its ability in many fields, like cross-language voice cloning~\cite{zhang2019learning} and singing voice conversion~\cite{nachmani2019unsupervised, deng2019pitchnet}.

With recent development of generative adversarial net (GAN)~\cite{goodfellow2014generative} in many tasks, such as image generation~\cite{reed2016generative, radford2015unsupervised}, text-to-speech~\cite{binkowski2019high, bollepalli2019generative} and others~\cite{pascual2017segan, yang2017midinet}, it has also been successfully applied to SVS field as a powerful generative model. Hono~\cite{hono2019singing} proposed a DNN-based SVS system with conditional GAN to optimize the distribution of acoustic features. Lee~\cite{lee2019adversarially} adopted a conditional adversarial training network in end-to-end SVS system. WGANSing~\cite{chandna2019wgansing} presented a block-wise generative singing model with Wasserstein-GAN~\cite{Martin2017WGAN} framework. In~\cite{Soonbeom2020Korean}, an auto-regressive singing model based on boundary equilibrium GAN was proposed. However, all the aforementioned GAN-based SVS systems only adopted a single discriminator directly operating on the whole sample sequences. They have a limitation, that is the lack of diversity in the sample distribution evaluation. Different from them, we introduce multiple random window discriminators (MRWDs)~\cite{binkowski2019high} into the multi-singer singing model to make the network be a GAN. It is an ensemble of discriminators operating on random sub-sampled fragments of samples. MWRDs allows for the evaluation of samples in different complementary ways, where it analyses the samples in the general realism, as well as in the correspondence between the generated samples and input conditions. Moreover, using different sizes of random windows, rather than the whole sample sequences, has a data augmentation effect.  It is quite helpful since the training data are limited for each singer.

%With a similar purpose, we introduce another variant of GAN network, multiple random window discriminators (MRWDs)~\cite{binkowski2019high} into the multi-singer singing model. Instead of a single discriminator in~\cite{hono2019singing, lee2019adversarially}, MRWDs is an ensemble of discriminators, which operates on random sub-sampled fragments of the real and generated samples. MRWDs analyses the audio in the general realism, as well as in the correspondence between the generated audio and input condition. %%Inspired by its significant performance in TTS task, we adopt MRWDs as adversarial training network in SVS system. 

In this paper, we propose a multi-singer sequence-to-sequence singing model based on adversarial training strategy. Our contributions includes: (1) Scale a Seq2Seq network to support multi-singer training. It improves the performance when only limited recordings are available for one singer. (2) Incorporate an adversarial task of singer classification to make encoder output less singer dependent to handle the data unbalance issue. (3) Apply multiple random window discriminators on the generated acoustic features to make the network be a GAN.

Similar to~\cite{blaauw2019sequence}, we only focus on spectrum modeling and assume $F0$ are given. Meanwhile, to avoid the impact of different duration models, ground-truth phoneme duration are used in both training and inference phases. WORLD vocoder~\cite{morise2016world} is used to extract acoustic features, which allows the explicit control of $F0$. In this way, we evaluate the contribution of each component and demonstrate our proposed singing model. %can obtain singing voice of higher quality and better pronunciation performance in high-pitched voice with limited training data.  

\section{Single-singer Seq2Seq SVS system}
%Recently, sequence-to-sequence (seq2seq) models have shown significant advantages in many tasks. Blaauw~\cite{blaauw2019sequence} proposed a sequence-to-sequence SVS system based on feed-forward transformer to show its strong ability in SVS task.
Following~\cite{blaauw2019sequence}, the system is shown in the solid lines of Figure~\ref{fig:struct}, including encoder, length regulator and decoder.

Firstly, the encoder takes phoneme embedding, pitch embedding and position encoding from musical score as input and then obtain its score encoding through a series of gated linear units (GLU) blocks~\cite{dauphin2017language}. GLU block is a convolutional block including a 1-D convolutional layer with gated linear units and a residual connection. The length regulator expands the score encoding from phoneme-level to frame-level according to phoneme duration. Finally, the feed-forward decoder transforms the expanded score encoding sequence to its corresponding acoustic feature sequence through several GLU blocks and self-attention layers~\cite{vaswani2017attention}. %In~\cite{blaauw2019sequence}, an approximate initial alignment is given by a simplistic phoneme duration model using heuristic algorithm. 

%Listening tests show that this Seq2Seq SVS system can achieve a better performance than auto-regressive model and it has a faster inference and can avoid exposure bias issues than auto-regressive model.

%Inspired by its significant performance in SVS task, we adopt it as our baseline system.  

\section{Proposed Architecture}
\begin{figure}[htb]
\begin{minipage}[b]{1.0\linewidth}
  \centering
  \centerline{\includegraphics[width=8.5cm]{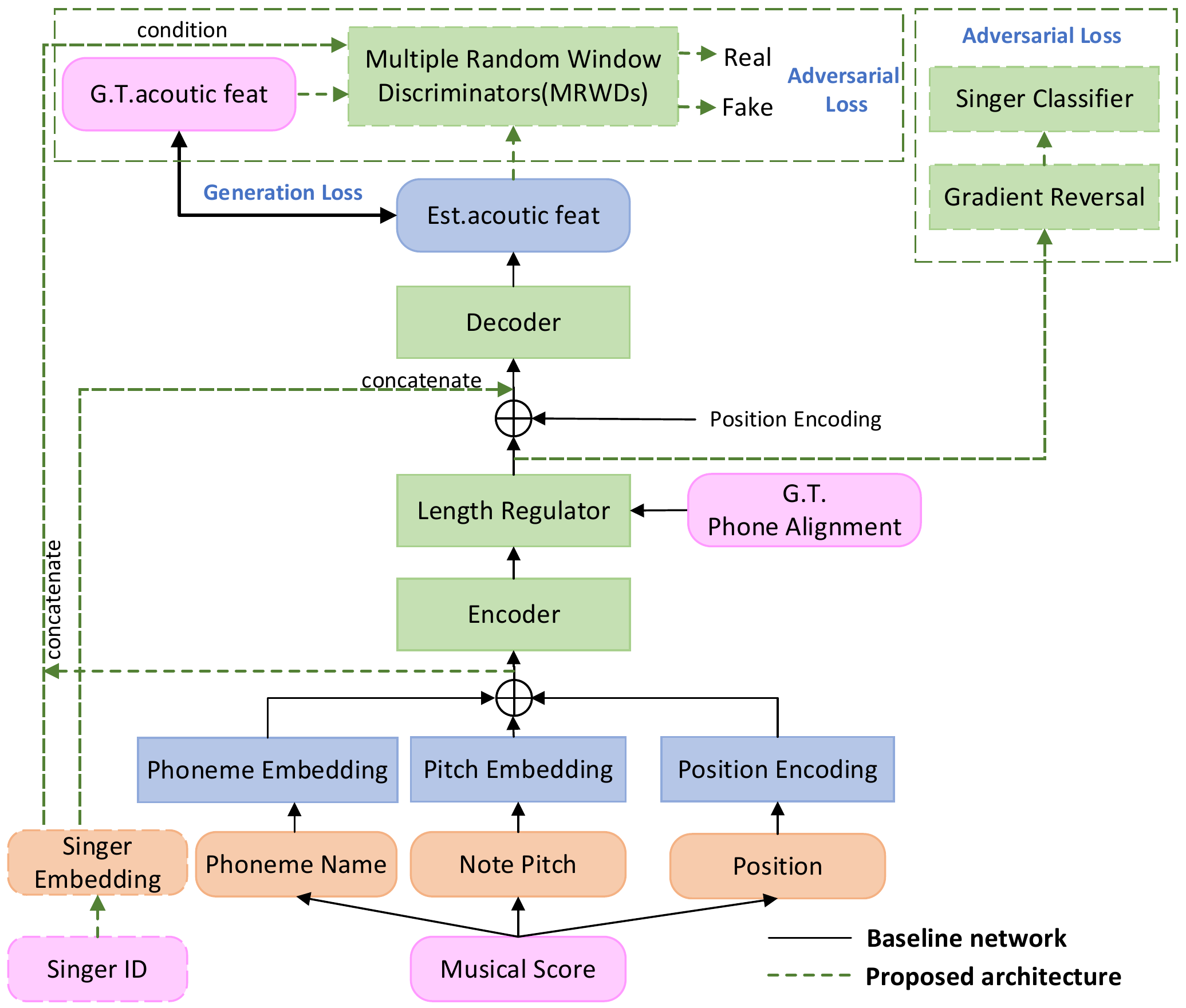}}
\end{minipage}
\vspace{-15pt}
\caption{Overview of the proposed adversarially trained multi-singer sequence-to-sequence singing synthesizer. Green dotted lines represents additional components in the proposed architecture while black solid lines is the baseline network.}
\vspace{-15pt}
\label{fig:struct}
\end{figure}

As illustrated in the dotted lines of Figure~\ref{fig:struct}, the proposed architecture consists of three modules: multi-singer SVS module, singer classifier and multiple random window discriminators. 

%Different from the single-singer model, the decoder is conditioned on the singer identity vector in the multi-singer model.%The multi-singer system is a seq2seq model based on feed-forward transformer, which is trained to generate acoustic features by conditioning on its corresponding singer identity.  Multiple random window discriminator is trained to make generated data distribution close to the true data distribution in an adversarial training manner. Especially, we employ domain adversarial training by introducing a singer classifier to encourage the encoder to learn disentangled representation of musical score and singer identity. It can effectively solve the data sparsity issue among multiple singers.
To focus on the spectrum and exclude the influence of duration model in prosody, in length regulator, we use ground-truth phoneme alignment in both training and inference phases. 

\begin{figure}[!t]
\begin{minipage}[b]{1.0\linewidth}
  \centering
  \centerline{\includegraphics[width=8cm]{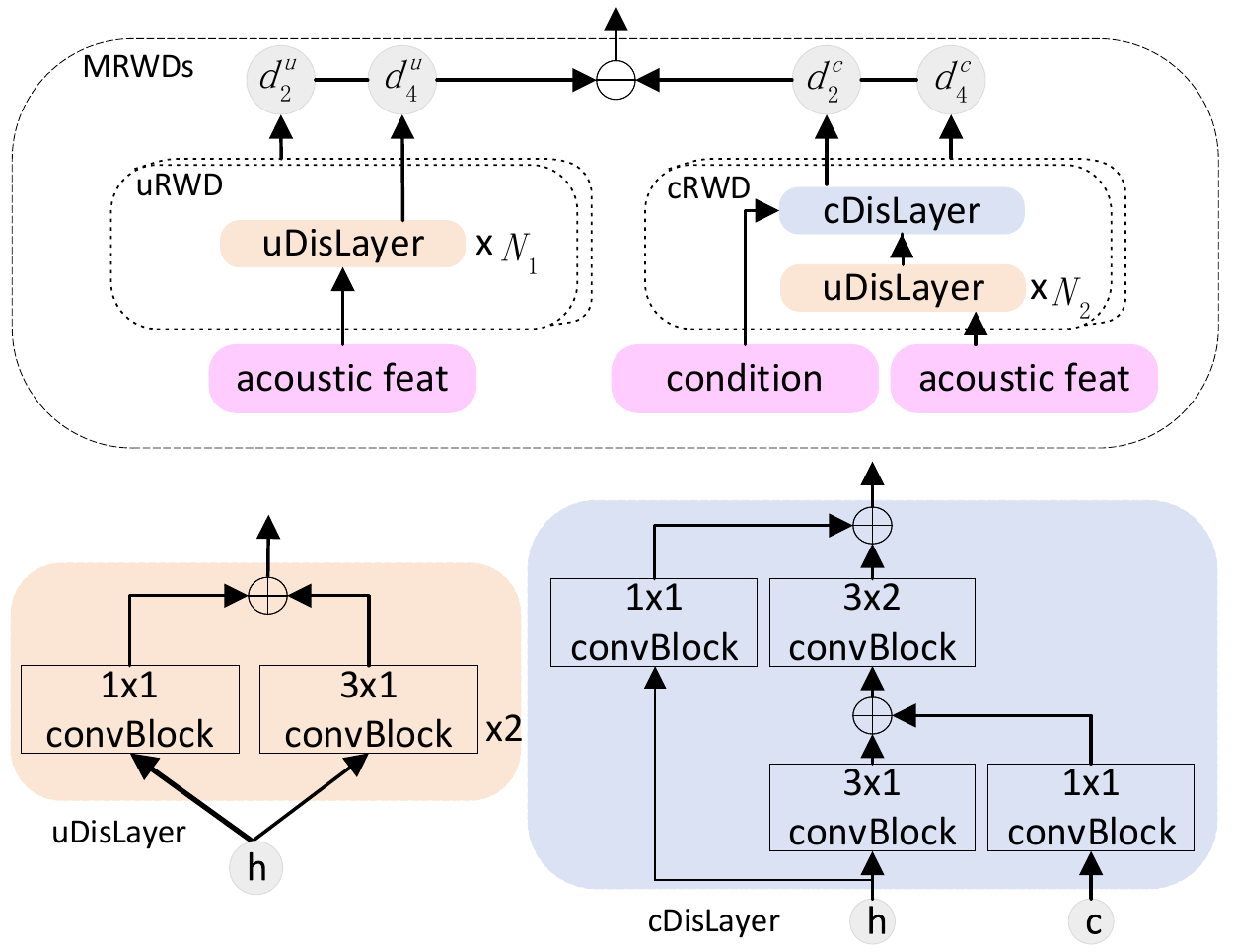}}
\end{minipage}
\vspace{-15pt}
\caption{The architecture of multiple random window discriminators (MRWDs). It contains 2 unconditional (uRWD, left) and 2 conditional (cRWD, right) discriminators with [2,4] random window size. The convBlock is a 1-D convolutional network with ReLU activation and spectral normalization~\cite{miyato2018spectral}.}
\vspace{-15pt}
\label{fig:MRWDs}
\end{figure}

\subsection{Multi-singer Seq2Seq SVS System}
For the purpose of reducing the training data size of target singer, one solution is to leverage singing data from other singers. For example, in~\cite{blaauw2019data}, a learned singer embedding from acoustic features was used to represent singer identity. An singer identity encoder was designed to produce singer's identity vector in SVS system~\cite{lee2019disentangling}. %Different from~\cite{blaauw2019data} and~\cite{lee2019disentangling}, to simplify the architecture of SVS system, we adopt a simple method which firstly generates an one-hot singer identity vector and then maps it to a continuous space. In the end, the continuous speaker embedding vector is concatenated with the encoder output in channel-wise as decoder input. 

Following~\cite{ping2017deep}, we construct our multi-singer singing model with trainable singer embedding. These singer embedding vectors are based on singer ID, which are initially randomized and then updated during training. It is concatenated with encoder output and frame position encoding as the input of decoder. %The input of decoder is the concatenation of the singer's embedding, the frame position encoding and score encoding at every time point. 

We use the cross-entropy loss function for voiced/unvoiced (VUV) flag prediction while L1 loss for mel-generalized coefficient
(MGC) and band aperiodicity (BAP) prediction. Therefore the generation loss for acoustic features is:
\vspace{-5pt}
\begin{align}\label{reconloss}
L_{G} = L_{1}^{mgc} + L_{1}^{bap} + L_{ce}^{vuv}
%\vspace{-25pt}
\end{align}

\subsection{Singer Classifier}
One challenge in multi-singer SVS system is the data unbalance issue among singers, which is also mentioned in cross-language voice cloning task~\cite{zhang2019learning}. To address this problem, we incorporate an adversarial loss by employing a singer classifier based on the musical score encoding. The encoder is expected to learn a latent representation which is strong on acoustic feature generation task while weak on singer classification. In other words, singer identity information is removed from encoder output.

As usual, a gradient reversal layer (GRL)~\cite{ganin2016domain} is applied prior to the singer classifier. It multiplies the gradient by a certain negative constant ($\lambda $) during back propagation of training while only acts an identity transformation during forward propagation.

For singer classifier, cross entropy loss function is applied as below:
\vspace{-5pt}
\begin{align}\label{scloss}
L_{adv_{singer}} = \sum_{i} L_{ce}(S(E(x_{i})), s_{l_{i}})
\end{align}
where $S$ is singer classifier and $E$ represents encoder, $i$ means the $i$-th singer and $s_{l}$ is the singer identity label. 

\subsection{Multiple Random Window Discriminators}
Expecting to generate more realistic singing voice, we adopt multiple random window discriminators (MWRDs)~\cite{binkowski2019high} to make the network be a GAN. %to make generated data distribution close to true data distribution. 

As shown in Figure~\ref{fig:MRWDs}, MRWDs contains unconditional and conditional discriminators with random window sizes (uRWD and cRWD). All these discriminators are independent of each other. It can measure the similarity of distribution between generated and real acoustic features as well as consider whether the generated samples match the input conditions (e.g., phoneme, pitch and singer identity). Furthermore, random window sizes in MRWDs allow different sub-segments feeding into these discriminators. On the one hand, using  different sizes of random window has a data augmentation effect. It is very helpful since the training data are limited for each singer. On the other hand, feeding sub-segments instead of the whole sample sequence can also reduce the computational complexity in training. Finally, outputs of all discriminators are added together as the final MRWDs output to calculate the loss. The final MRWDs output is:
\vspace{-10pt}
\begin{align*}
MRWDs_{(2,4)} = \Sigma_{k\in(2,4)}(uRWD_{k} + cRWD_{k}) 
\end{align*}

Following the vanilla GAN loss in~\cite{goodfellow2014generative}, the adversarial loss terms ($L_{adv_{D}}$ and $L_{adv_{G}}$) for MRWDs $D$ and generator $G$ are shown below:
\vspace{-5pt}
\begin{align}\label{eq:GANLoss}
\begin{split}
L_{adv_{D}} = & -\mathbb{E}_{y\sim P_{r}}\left [ log(D(y)) \right ]\\
              & -\mathbb{E}_{x\sim P_{x}}\left [ log(1 - D(G(x)) \right ]
\end{split}\\
L_{adv_{G}} = &  \mathbb{E}_{x\sim P_{x}}\left [ log(1 - D(G(x)) \right ]
\end{align}
where $y$ is a sample from real data distribution and $x$ is the input of the generator. Different from the vanilla GAN~\cite{goodfellow2014generative}, the input $x$ is musical score feature instead of random noise.

\section{Experiments}

\subsection{Dataset}
\begin{itemize}
    \item {\textbf{Single-singer SVS system}}: 770 Chinese pop songs (total $ \sim $10 hours) of clean recordings are collected from a female singer ``F1''.
    \item {\textbf{Multi-singer SVS system}}:  200 songs ($ \sim $3.5 hours) of ``F1'' are randomly selected from 770 songs. Meanwhile, about 200 Chinese pop songs are also collected from six singers, named ``F2'', ``F3'', ``F4'', ``M1'', ``M2'' and ``M3'' respectively\footnote{ Here, ``F'' means female singer while ``M'' means male singer.}. The specific number of songs from each singer are listed in Table~\ref{tab:songsnum}.
\end{itemize}

\vspace{-10pt}
\begin{table}[h]
\scriptsize
\renewcommand{\arraystretch}{1.6}
\addtolength{\tabcolsep}{0pt}
\caption{The number of songs from seven singers.}\label{tab:songsnum}
\vspace{-10pt}
\centering
\begin{tabular}{|c|c|c|c|c|c|c|c|}
\cline{1-8}
\hline
                & F1  & F2  & F3  & F4  & M1 & M2 & M3 \\
\hline
Number of songs & 200 & 200 & 200 & 210 &205 &200 & 153\\
\hline
\end{tabular}
\vspace{-5pt}
\end{table}

All the recordings are collected in a professional recording studio while singers listen to the accompaniment through headphones. The musical scores are hereafter manually revised according to the audios. The phoneme duration is obtained by HMM-based force alignment tool~\cite{sjolander2003hmm}.

Recordings are sampled at 48kHz with 16bits per sample in mono. They are segmented into pieces at silence boundaries. Each piece is forced to be not longer than 10 seconds. Acoustic features are then extracted from audio pieces at every 15ms by WORLD vocoder, including 60-dimensional MGC, 5-dimensional BAP and 1-dimensional VUV flag.

\subsection{Experimental setup}
In the experiments, lyrics and note pitch in MIDI standard format~\cite{midi} are obtained from musical score files. %Phoneme sequence is generated by text analysis tool. %text analysis is used to get phoneme sequence from lyric text. There exists three phonemes in one Chinese character, one for initial and two for final. Meanwhile, we can obtain note pitch in  files. 
The sizes of phoneme and note pitch vocabularies are 71 and 84 respectively. 
As shown in Figure~\ref{fig:struct}, phoneme and note pitch are both embedded into 384-dimensional vectors and added together with position encoding as encoder input. %384-dim input musical score vector are generated from 384-dim phoneme embedding, pitch embedding and position encoding vectors. %phoneme and pitch are embedded to 384-dim vector individually and then added together with 384-dim position encoding vector to generate the 384-dim input musical score sequence. 64-dim singer embedding vectors are obtained from 7-dim one-hot singer identity vector for all singers. 
Besides, additional 64-dimensional trainable singer embedding vector is used to represent singer identity .

\begin{figure}[ht]
\begin{minipage}[b]{1.0\linewidth}
  \centering
  \centerline{\includegraphics[width=6cm]{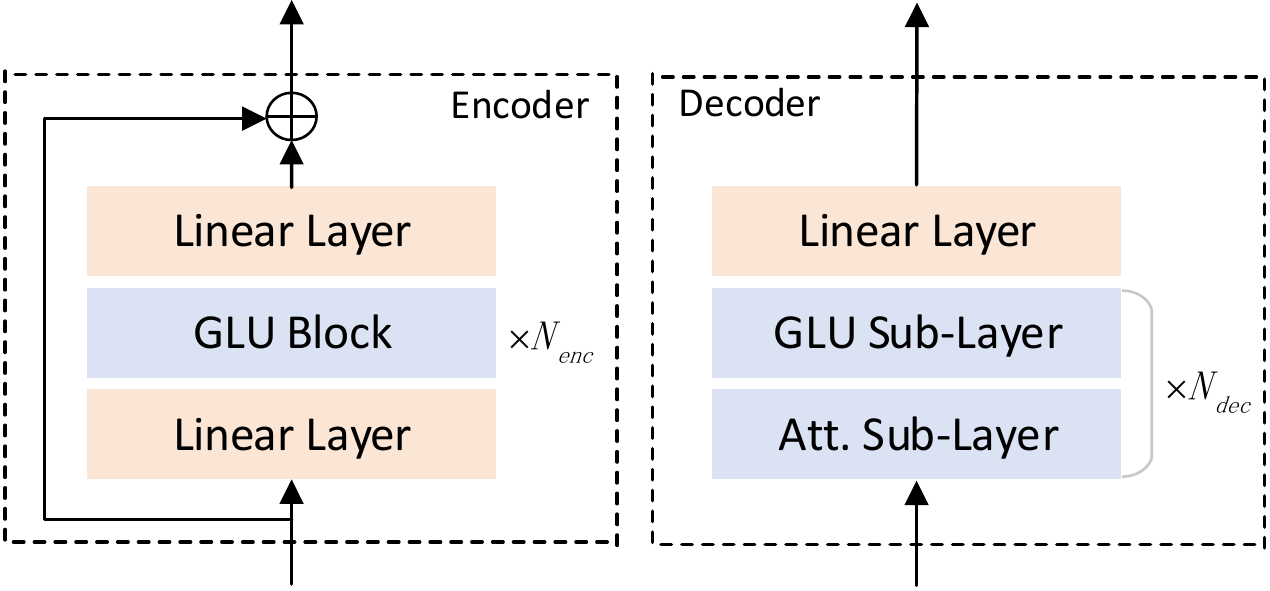}}
\end{minipage}
\vspace{-15pt}
\caption{A diagram of the encoder and decoder networks, where GLU block, GLU sub-layer and attention sub-layer follows the design of~\cite{blaauw2019sequence}.}
\vspace{-15pt}
\label{fig:encoderdecoder}
\end{figure}

Following the design in~\cite{blaauw2019sequence}, the networks of encoder and decoder are shown in Figure~\ref{fig:encoderdecoder}. The encoder contains two linear layers and a single GLU block. The input/output size of each layer is 384/256, 256/64 and 64/384 respectively. Decoder is a 6-layers network. Each layer consists of a single-head self-attention sub-layer and a GLU sub-layer, all with 448 channels. The output linear layer maps the 448-dimensional hidden vector to 66-dimensional acoustic feature vector. The kernel sizes of convolutional layers in encoder and decoder are both 3. 
%\vspace{-15pt}

The singer classifier is a 2-layers 1-D convolutional network with ReLU activation and spectral normalization~\cite{miyato2018spectral}. The kernel sizes are all set to be 3, with input/output size of 384/128 for the first layer and 128/128 for the second layer. The output linear layer finally converts the 128-dimensional hidden vector into a 7-dimensional singer identity vector with softmax activation.

As shown in the top of Figure~\ref{fig:MRWDs}, MRWDs consists of 2 uRWDs and 2 cRWDs with input window size of [2,4] in frames respectively. uRWD is a stack of 4 uDisLayers while cRWD contains 3 uDisLayers and 1 cDisLayer. The networks of uDisLayer and  cDisLayer are shown in the bottom of Figure~\ref{fig:MRWDs}, which are both a stack of convolutional networks with different kernel sizes. uRWD and cRWD have the same output channels of [64, 128, 256, 1] for 4 layers. %In addition, in cWRDs, we concatenate 64-dim singer embedding with 384-dim input musical score feature to form 448-dim condition vector. 
In cRWDs, the input condition is a 448-dimensional vector which is the concatenation of 384-dimensional score encoding and 64-dimensional singer embedding.  Finally, all output of these 4 discriminators are added together as the final MRWDs output to calculate loss. 

%When applying MRWDs as discriminator, the seq2seq SVS model is also namely generator, the entire model is trained with a single discriminator step per generator step sharing the initial learning rate of $1 \times 10^{-4}$.
In the training of GAN, %the singing model performs one step discriminator update between each single step of generator update.  
discriminator and generator are updated alternately sharing the initial learning rate of $1 \times 10^{-4}$. The mini batch size is 32 and the Adam optimizer is with $\beta _{1} = 0.9, \beta _{2} = 0.98$. Dropout probability is 0.1 through the entire model. 

We implement five Seq2Seq SVS systems to evaluate the contributions of three proposed modules. The differences among the five systems are described in Table~\ref{tab:systems}. The baseline and three modules are:
\begin{itemize}
    \item {\textbf{baseline}}: Single-singer singing model.
    \item {\textbf{module1}}: Multi-singer singing model. 
    \item {\textbf{module2}}: Singer classifier.
    \item {\textbf{module3}}: Multiple random window discriminators. 
\end{itemize}

In the training phase, the total generative loss is represented as follows:
\begin{align}\label{eq:totalloss}
L_{total} = \lambda _{G}L_{G}  +  \lambda _{S}L_{adv_{singer}} + \lambda _{D}L_{adv_{G}}
\end{align}
where $\lambda _{G}$, $\lambda _{S}$ and $\lambda _{D}$ are the weights for the losses of the three modules. 

Five systems have different weights for each module, which are specifically listed in Table~\ref{tab:systems}.
%Modules and their corresponding loss weights [$\lambda _{recon}$, $\lambda _{D}$, $\lambda _{S}$] in these five SVS systems are described in Table~\ref{tab:systems}
\vspace{-5pt}
\begin{table}[ht]
\scriptsize
\renewcommand{\arraystretch}{1.5}
\addtolength{\tabcolsep}{-5pt}
\caption{Five systems description of modules and loss weights.}\label{tab:systems}
\vspace{-10pt}
\centering
\begin{tabular}{|c|c|c|c|c|c|}
\cline{1-6}
\hline
Systems  &System1  & System2    & System3    &System4       &System5\\
\hline
Modules  &baseline & +(module1) & +(module1,2) &+(module1,3) &+(module1,2,3)\\
\hline
Loss weights  &[1,0,0] &[1,0,0] & [1,1,0] &[10,0,1] &[10,2,1]\\
\hline
\end{tabular}
\end{table}
\vspace{-10pt}

\subsection{Evaluation}
To evaluate the performance of different systems, we conduct objective and subjective evaluations on the target singer ``F1''. 
\subsubsection{Objective Evaluation}
\begin{figure}[ht]
\begin{minipage}[b]{1.0\linewidth}
  \centering
  \centerline{\includegraphics[width=8cm]{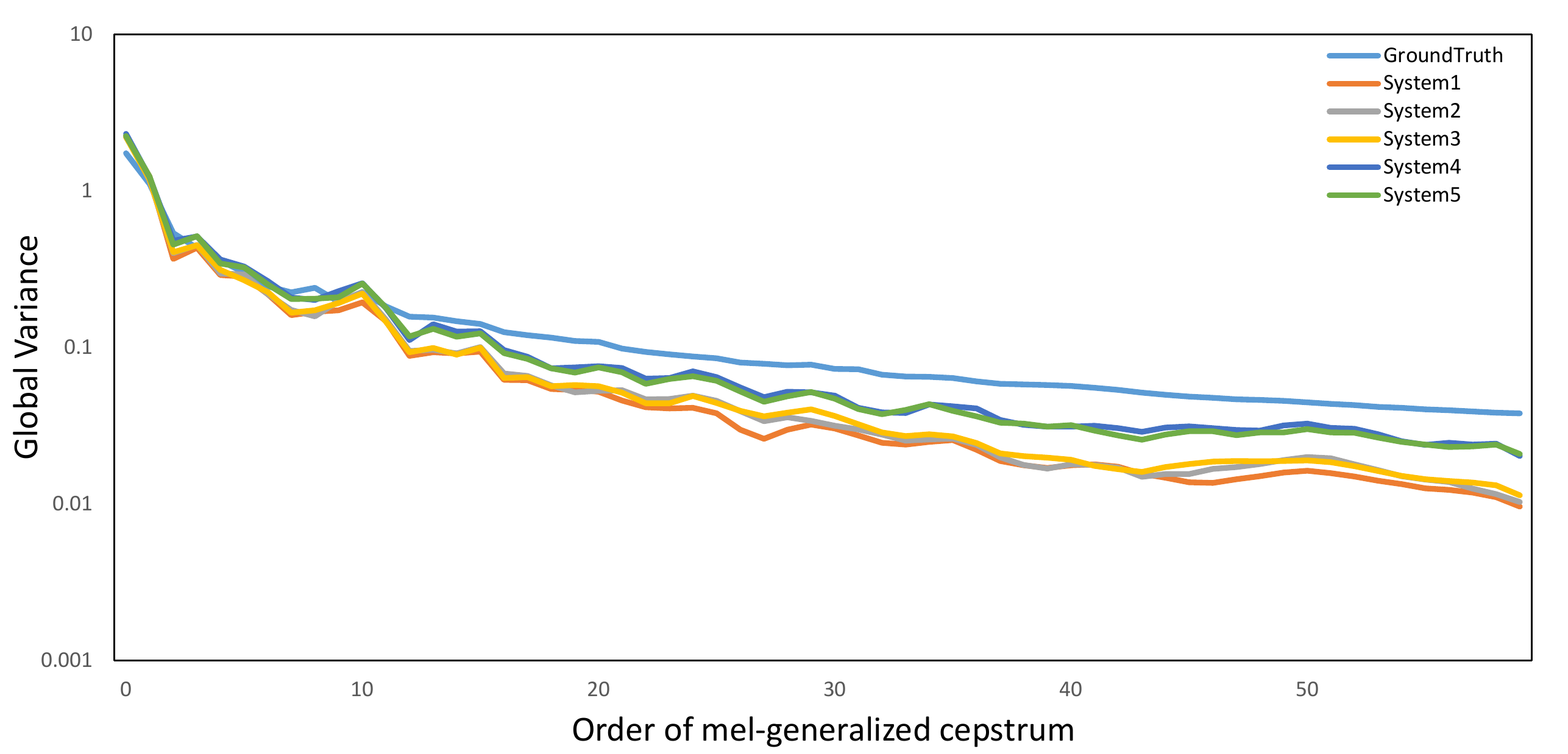}}
\end{minipage}
\vspace{-15pt}
\caption{Averaged global variances (GVs) of mel-generalized coefficient (MGC) from  evaluation set for different systems.}
\label{fig:globalvariance}
\end{figure}

In objective test, averaged global variances (GVs)~\cite{toda2007voice} of MGC is calculated for different systems. As shown in Figure~\ref{fig:globalvariance}, system2 and system3 have similar averaged GVs and both slightly better than System1. It indicates that multi-singer module can even improve the averaged GVs with less training data of target singer. Furthermore, the GVs trajectories of system4 and system5 are on par but much closer to ground truth than others. It shows that MRWDs can significantly alleviated the over-smoothing problem of generated features. Also, it can be noted that singer classifier has no obvious impact to GVs.
%\vspace{-5pt}
\vspace{-5pt}

\begin{figure}[ht]
\begin{minipage}[b]{1.0\linewidth}
  \centering
  \centerline{\includegraphics[width=8.5cm]{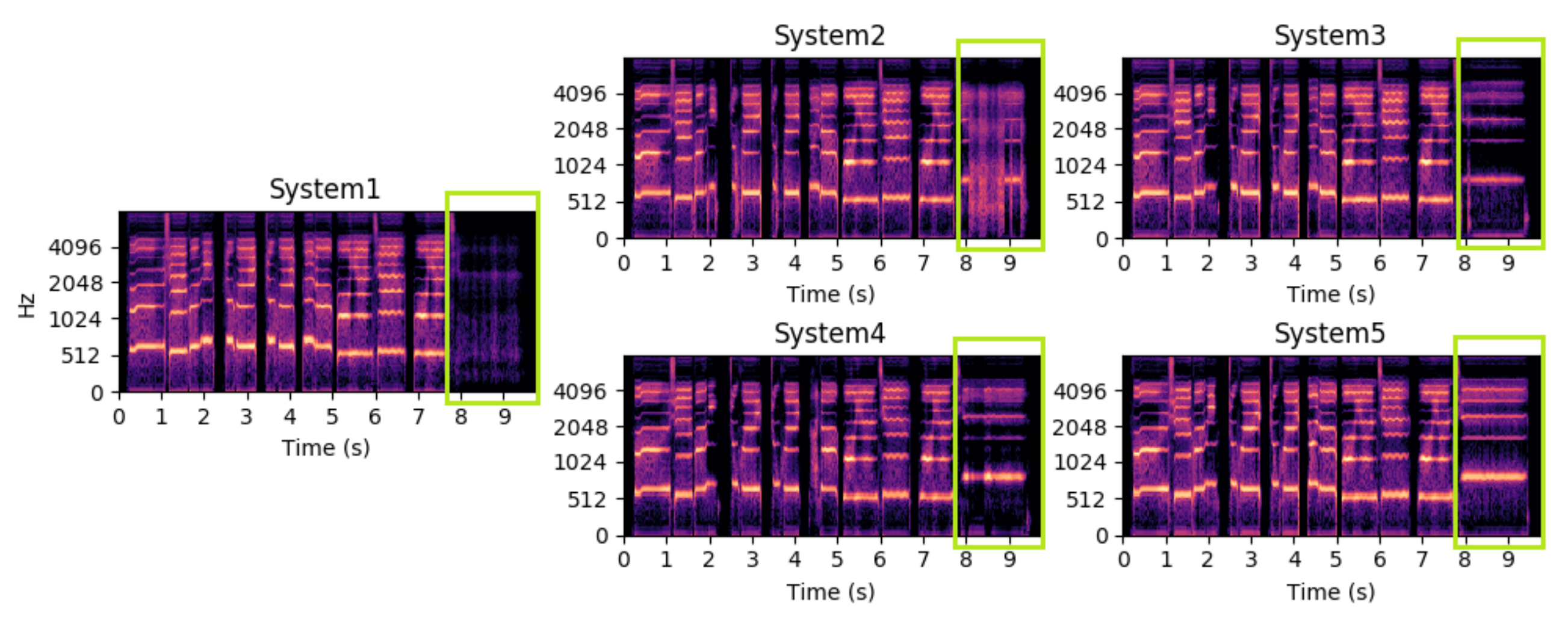}}
\end{minipage}
\vspace{-15pt}
\caption{Mel-spectrograms of a testing sample generated by different systems. Green rectangles show their differences of the same character.}
\vspace{-15pt}
\label{fig:mel-spectrogram}
\end{figure}

We visualize the mel-spectrograms of a testing sample generated by different systems in Figure~\ref{fig:mel-spectrogram}. These green rectangles refer to a character ``qi a\_h nn\_h'' whose note pitch is 80 in MIDI standard. As shown in the green rectangles, the mel-sepctrogram of the character in System1 is almost invisible while in System2 it has normal energy although quite fuzzy harmonics. This phenomenon can be attribute to the benefit of leveraging data from other singers. Based on it, both System3 and System4 show correct and clear harmonics against System2. It indicates that either singer classifier or MRWDs can obviously enhance the articulation of high-pitched vowels. Moreover, when combining singer classifier and MRWDs together, System5 achieves further improvement and defeat all other systems. Overall, from System1 to System5, it can be easily found that the proposed synthesizer enhance the articulation of high-pitched vowels step by step. %Particularly, from the two comparative results of System2 with System3 and System4 with System5, we also note that incorporating a singer classifier can effectively improve the articulation in high-pitched vowels whatever MRWDs is employed.%In the upper of the figure, we can see that system3 obtains a brighter and more clear frequency band than system2 in the green rectangular. Similar comparison results can obtain between system4 and system5 in the bottom of the figure. It show that when using limited singing data in multi-singer system, the data unbalance issue might result in the poor articulation performance, especially in the high pitch band while singer classifier can significantly improve the articulation performance to solve the data sparsity issue. Later we will confirm our conclusion in the subjective evaluation.   
\vspace{-10pt} 

\subsubsection{Subjective Evaluation}
For subjective evaluation, the female singer ``F1'' is regarded as target singer. WORLD vocoder is used to synthesize the singing voice with generated MGC, BAP and ground-truth $F0$. 

Firstly, Mean Opinion Score (MOS) test is carried out to evaluate the quality of generated voices from different systems. 10 listeners are invited to judge 30 samples in each system. Each sample is not longer than 10 seconds.
%The results of MOS scores for singing voice quality are shown in Figure~\ref{fig:mos}.

\begin{figure}[ht]
\begin{minipage}[b]{1.0\linewidth}
  \centering
  \centerline{\includegraphics[width=8cm]{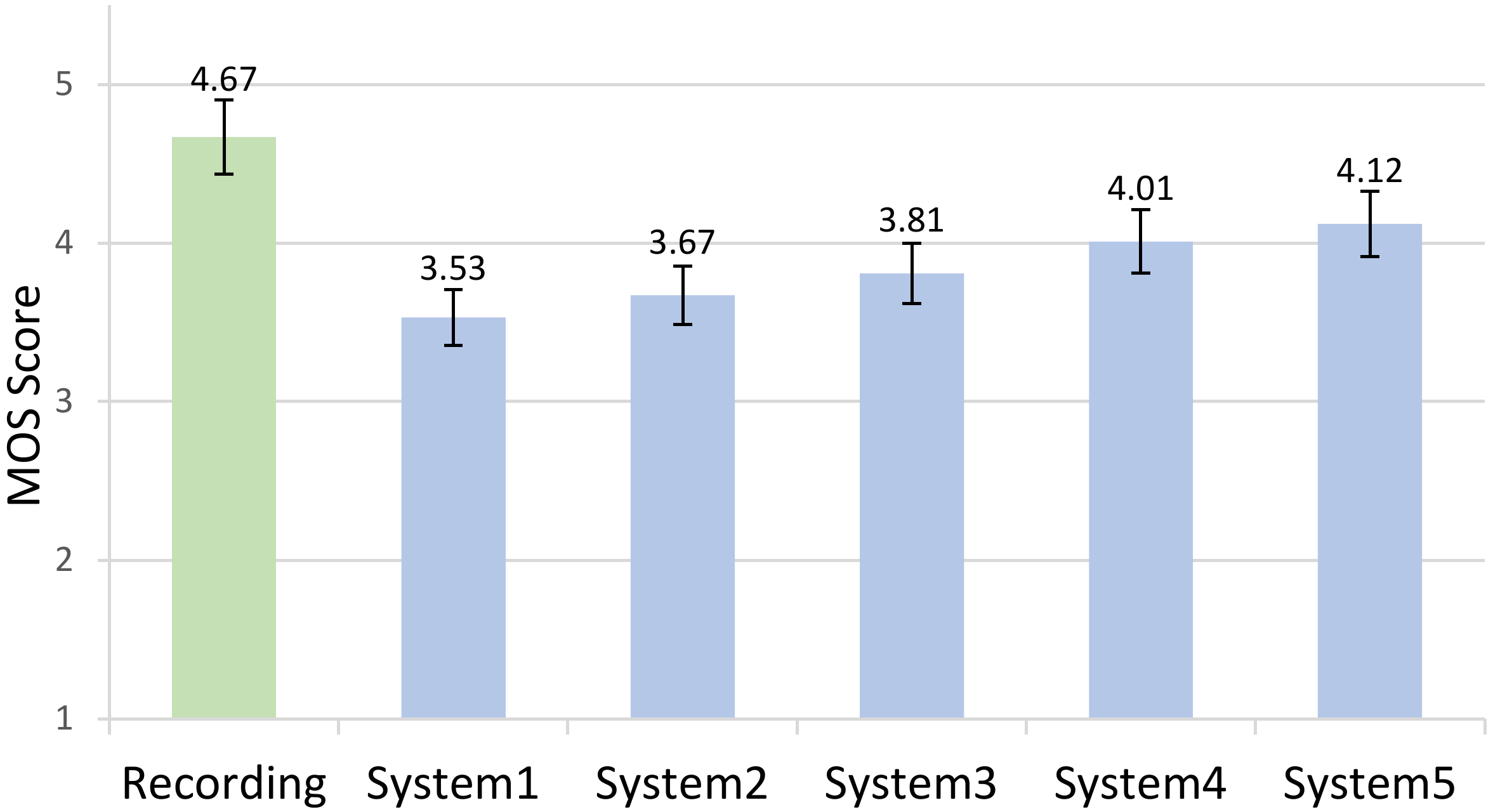}}
\end{minipage}
\vspace{-15pt}
\caption{ Mean Opinion Score (MOS) test results of singing voice quality with $95\%$ confidence intervals.}
%\vspace{-5pt}
\label{fig:mos}
\end{figure}

As shown in Figure~\ref{fig:mos}, System2 achieves a higher MOS score than System1, proving the benefit of leveraging data of other singers. Multi-singer singing model can improve the quality even with less singing data of target singer. Meanwhile, the MOS scores of System3 and System4 are both higher than that of System2, confirming the effectiveness of integrating singer classifier or MRWDs individually. Our proposed System5 achieves the highest score of 4.12. It demonstrates that further advancement can be achieved if combining singer classifier and MRWDs together. % showing that merging singer classifier and MRWDs together into multi-singer singing model obtains the best performance in singing voice quality. 

Particularly, two A/B preference tests are conducted to validate the effect of singer classifier: System2 vs. System3 and System4 vs. System5. For each system, 30 samples are generated and each one contains 1-5 vowels whose pitch ranges from 78 to 81 in MIDI standard. 10 listeners are required to listen to all the pairs and choose which one is better on the articulation. %P-values are used to determine the significance of the results. The smaller the p-values, the larger the significance.

\begin{figure}[ht]
\begin{minipage}[b]{1.0\linewidth}
  \centering
  \centerline{\includegraphics[width=8cm]{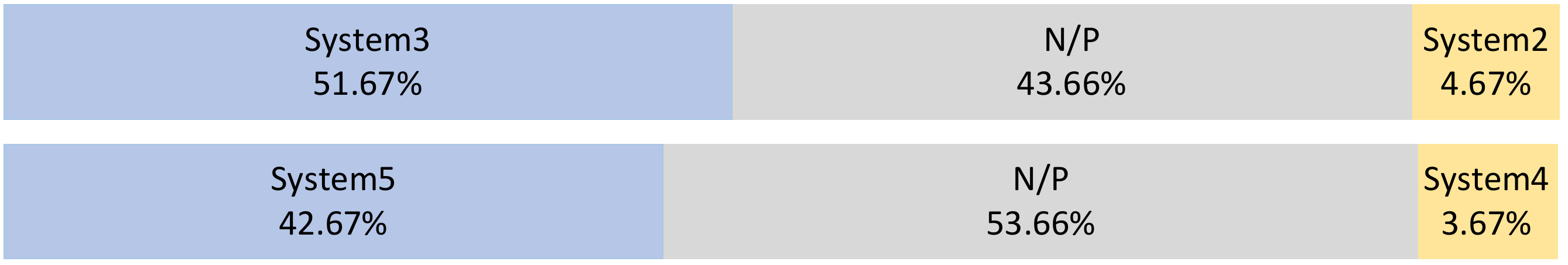}}
\end{minipage}
\vspace{-15pt}
\caption{A/B preference tests on the articulation of high-pitched vowels between different systems. The p-values for the two bars are $1.14*10^{-9}$ and $2.17*10^{-9}$ respectively.}
\label{fig:cmos}
\end{figure}

The results are shown in Figure~\ref{fig:cmos}. Both A/B tests prove the overwhelming advantage of systems with adversarial task of singer classification. They are mainly contributed by the model quality improvement on sparse input (such as high pitch).  It is likely because the model of one singer can better leverage others' data when the distribution of encoder output is less singer dependent.
%From the upper bar in Figure~\ref{fig:cmos}, it can be found that System3 achieves 51.67\% preference rate, which is significantly higher than 4.67\% of System2. The bottom bar in Figure~\ref{fig:cmos} shows that 42.67\% supports System5 while only 3.67\% supports System4. These results show that singer classifier has an important impact whatever MRWDs is employed. The singer classifier can effectively improve the articulation naturalness in high-pitched vowels which is consistent with the conclusion of Figure~\ref{fig:mel-spectrogram}.

\vspace{-10pt}

\vspace{-15pt}
\vspace{-15pt}

\section{Conclusions}

In this paper, we extend conventional  sequence-to-sequence singing synthesizer to a multi-singer one with trainable singer embedding which can model each singer well with limited recording data. Moreover, adversarial training strategies are introduced into the multi-singer synthesizer. One is the singer classifier to exclude singer identity information from encoder output. The other is MRWDs for acoustic feature generation loss to help the synthesizer to generate more realistic singing voice. Experimental results prove that our methods can achieve higher quality singing voice and especially significant improvement on the articulation of high-pitched vowels. By this method, only limited recordings are required from target singer for a high quality singing model. In other words, the singing voice customization becomes easy and economic. Some samples for subjective evaluation are available via this link\footnote{https://jiewu-demo.github.io/INTERSPEECH2020/}.

%\section{Acknowledgements}

%The ISCA Board would like to thank the organizing committees of the past INTERSPEECH conferences for their help and for kindly providing the template files. \\
%Note to authors: Authors should not use logos in acknowledgement section; rather authors should acknowledge corporations by naming them only.

\input{reference.bbl}

\bibliographystyle{IEEEtran}

\bibliography{reference}

% \begin{thebibliography}{9}
% \bibitem[1]{Davis80-COP}
%   S.\ B.\ Davis and P.\ Mermelstein,
%   ``Comparison of parametric representation for monosyllabic word recognition in continuously spoken sentences,''
%   \textit{IEEE Transactions on Acoustics, Speech and Signal Processing}, vol.~28, no.~4, pp.~357--366, 1980.
% \bibitem[2]{Rabiner89-ATO}
%   L.\ R.\ Rabiner,
%   ``A tutorial on hidden Markov models and selected applications in speech recognition,''
%   \textit{Proceedings of the IEEE}, vol.~77, no.~2, pp.~257-286, 1989.
% \bibitem[3]{Hastie09-TEO}
%   T.\ Hastie, R.\ Tibshirani, and J.\ Friedman,
%   \textit{The Elements of Statistical Learning -- Data Mining, Inference, and Prediction}.
%   New York: Springer, 2009.
% \bibitem[4]{YourName17-XXX}
%   F.\ Lastname1, F.\ Lastname2, and F.\ Lastname3,
%   ``Title of your INTERSPEECH 2020 publication,''
%   in \textit{Interspeech 2020 -- 20\textsuperscript{th} Annual Conference of the International Speech Communication Association, September 15-19, Graz, Austria, Proceedings, Proceedings}, 2020, pp.~100--104.
% \end{thebibliography}

\end{document}

%% file: reference.bbl

%% file: template.bbl
\begin{thebibliography}{10}
\providecommand{\url}[1]{#1}
\csname url@samestyle\endcsname
\providecommand{\newblock}{\relax}
\providecommand{\bibinfo}[2]{#2}
\providecommand{\BIBentrySTDinterwordspacing}{\spaceskip=0pt\relax}
\providecommand{\BIBentryALTinterwordstretchfactor}{4}
\providecommand{\BIBentryALTinterwordspacing}{\spaceskip=\fontdimen2\font plus
\BIBentryALTinterwordstretchfactor\fontdimen3\font minus
  \fontdimen4\font\relax}
\providecommand{\BIBforeignlanguage}[2]{{%
\expandafter\ifx\csname l@#1\endcsname\relax
\typeout{** WARNING: IEEEtran.bst: No hyphenation pattern has been}%
\typeout{** loaded for the language `#1'. Using the pattern for}%
\typeout{** the default language instead.}%
\else
\language=\csname l@#1\endcsname
\fi
#2}}
\providecommand{\BIBdecl}{\relax}
\BIBdecl

\bibitem{nishimura2016singing}
M.~Nishimura, K.~Hashimoto, K.~Oura, Y.~Nankaku, and K.~Tokuda, ``Singing voice
  synthesis based on deep neural networks.'' in \emph{Interspeech}, 2016, pp.
  2478--2482.

\bibitem{kim2018korean}
J.~Kim, H.~Choi, J.~Park, S.~Kim, J.~Kim, and M.~Hahn, ``Korean singing voice
  synthesis system based on an lstm recurrent neural network,'' in \emph{Proc.
  INTERSPEECH}, 2018.

\bibitem{shen2018natural}
J.~Shen, R.~Pang, R.~J. Weiss, M.~Schuster, N.~Jaitly, Z.~Yang, Z.~Chen,
  Y.~Zhang, Y.~Wang, R.~Skerrv-Ryan \emph{et~al.}, ``Natural tts synthesis by
  conditioning wavenet on mel spectrogram predictions,'' in \emph{2018 IEEE
  International Conference on Acoustics, Speech and Signal Processing
  (ICASSP)}.\hskip 1em plus 0.5em minus 0.4em\relax IEEE, 2018, pp. 4779--4783.

\bibitem{blaauw2017neural}
M.~Blaauw and J.~Bonada, ``A neural parametric singing synthesizer modeling
  timbre and expression from natural songs,'' \emph{Applied Sciences}, vol.~7,
  no.~12, p. 1313, 2017.

\bibitem{yi2019singing}
Y.-H. Yi, Y.~Ai, Z.-H. Ling, and L.-R. Dai, ``Singing voice synthesis using
  deep autoregressive neural networks for acoustic modeling,'' \emph{arXiv
  preprint arXiv:1906.08977}, 2019.

\bibitem{valle2019mellotron}
R.~Valle, J.~Li, R.~Prenger, and B.~Catanzaro, ``Mellotron: Multispeaker
  expressive voice synthesis by conditioning on rhythm, pitch and global style
  tokens,'' \emph{arXiv preprint arXiv:1910.11997}, 2019.

\bibitem{blaauw2019sequence}
M.~Blaauw and J.~Bonada, ``Sequence-to-sequence singing synthesis using the
  feed-forward transformer,'' \emph{arXiv preprint arXiv:1910.09989}, 2019.

\bibitem{ren2019fastspeech}
Y.~Ren, Y.~Ruan, X.~Tan, T.~Qin, S.~Zhao, Z.~Zhao, and T.-Y. Liu, ``Fastspeech:
  Fast, robust and controllable text to speech,'' in \emph{Advances in Neural
  Information Processing Systems}, 2019, pp. 3165--3174.

\bibitem{ganin2016domain}
Y.~Ganin, E.~Ustinova, H.~Ajakan, P.~Germain, H.~Larochelle, F.~Laviolette,
  M.~Marchand, and V.~Lempitsky, ``Domain-adversarial training of neural
  networks,'' \emph{The Journal of Machine Learning Research}, vol.~17, no.~1,
  pp. 2096--2030, 2016.

\bibitem{zhang2019learning}
Y.~Zhang, R.~J. Weiss, H.~Zen, Y.~Wu, Z.~Chen, R.~Skerry-Ryan, Y.~Jia,
  A.~Rosenberg, and B.~Ramabhadran, ``Learning to speak fluently in a foreign
  language: Multilingual speech synthesis and cross-language voice cloning,''
  \emph{arXiv preprint arXiv:1907.04448}, 2019.

\bibitem{nachmani2019unsupervised}
E.~Nachmani and L.~Wolf, ``Unsupervised singing voice conversion,'' \emph{arXiv
  preprint arXiv:1904.06590}, 2019.

\bibitem{deng2019pitchnet}
C.~Deng, C.~Yu, H.~Lu, C.~Weng, and D.~Yu, ``Pitchnet: Unsupervised singing
  voice conversion with pitch adversarial network,'' \emph{arXiv preprint
  arXiv:1912.01852}, 2019.

\bibitem{goodfellow2014generative}
I.~Goodfellow, J.~Pouget-Abadie, M.~Mirza, B.~Xu, D.~Warde-Farley, S.~Ozair,
  A.~Courville, and Y.~Bengio, ``Generative adversarial nets,'' in
  \emph{Advances in neural information processing systems}, 2014, pp.
  2672--2680.

\bibitem{reed2016generative}
S.~Reed, Z.~Akata, X.~Yan, L.~Logeswaran, B.~Schiele, and H.~Lee, ``Generative
  adversarial text to image synthesis,'' \emph{arXiv preprint
  arXiv:1605.05396}, 2016.

\bibitem{radford2015unsupervised}
A.~Radford, L.~Metz, and S.~Chintala, ``Unsupervised representation learning
  with deep convolutional generative adversarial networks,'' \emph{arXiv
  preprint arXiv:1511.06434}, 2015.

\bibitem{binkowski2019high}
M.~Bi{\'n}kowski, J.~Donahue, S.~Dieleman, A.~Clark, E.~Elsen, N.~Casagrande,
  L.~C. Cobo, and K.~Simonyan, ``High fidelity speech synthesis with
  adversarial networks,'' \emph{arXiv preprint arXiv:1909.11646}, 2019.

\bibitem{bollepalli2019generative}
B.~Bollepalli, L.~Juvela, and P.~Alku, ``Generative adversarial network-based
  glottal waveform model for statistical parametric speech synthesis,''
  \emph{arXiv preprint arXiv:1903.05955}, 2019.

\bibitem{pascual2017segan}
S.~Pascual, A.~Bonafonte, and J.~Serra, ``Segan: Speech enhancement generative
  adversarial network,'' \emph{arXiv preprint arXiv:1703.09452}, 2017.

\bibitem{yang2017midinet}
L.-C. Yang, S.-Y. Chou, and Y.-H. Yang, ``Midinet: A convolutional generative
  adversarial network for symbolic-domain music generation,'' \emph{arXiv
  preprint arXiv:1703.10847}, 2017.

\bibitem{hono2019singing}
Y.~Hono, K.~Hashimoto, K.~Oura, Y.~Nankaku, and K.~Tokuda, ``Singing voice
  synthesis based on generative adversarial networks,'' in \emph{ICASSP
  2019-2019 IEEE International Conference on Acoustics, Speech and Signal
  Processing (ICASSP)}.\hskip 1em plus 0.5em minus 0.4em\relax IEEE, 2019, pp.
  6955--6959.

\bibitem{lee2019adversarially}
J.~Lee, H.-S. Choi, C.-B. Jeon, J.~Koo, and K.~Lee, ``Adversarially trained
  end-to-end korean singing voice synthesis system,'' \emph{arXiv preprint
  arXiv:1908.01919}, 2019.

\bibitem{chandna2019wgansing}
P.~Chandna, M.~Blaauw, J.~Bonada, and E.~G{\'o}mez, ``Wgansing: A multi-voice
  singing voice synthesizer based on the wasserstein-gan,'' in \emph{2019 27th
  European Signal Processing Conference (EUSIPCO)}.\hskip 1em plus 0.5em minus
  0.4em\relax IEEE, 2019, pp. 1--5.

\bibitem{Martin2017WGAN}
M.~Arjovsky, S.~Chintala, and L.~Bottou, ``{W}asserstein generative adversarial
  networks,'' in \emph{Proceedings of the 34th International Conference on
  Machine Learning}, 2017, pp. 214--223.

\bibitem{Soonbeom2020Korean}
S.~Choi, W.~Kim, S.~Park, S.~Yong, and J.~Nam, ``Korean singing voice synthesis
  based on auto-regressive boundary equilibrium gan,'' in \emph{ICASSP 2020},
  2020, pp. 7234--7238.

\bibitem{morise2016world}
M.~Morise, F.~Yokomori, and K.~Ozawa, ``World: a vocoder-based high-quality
  speech synthesis system for real-time applications,'' \emph{IEICE
  TRANSACTIONS on Information and Systems}, vol.~99, no.~7, pp. 1877--1884,
  2016.

\bibitem{dauphin2017language}
Y.~N. Dauphin, A.~Fan, M.~Auli, and D.~Grangier, ``Language modeling with gated
  convolutional networks,'' in \emph{Proceedings of the 34th International
  Conference on Machine Learning-Volume 70}.\hskip 1em plus 0.5em minus
  0.4em\relax JMLR. org, 2017, pp. 933--941.

\bibitem{vaswani2017attention}
A.~Vaswani, N.~Shazeer, N.~Parmar, J.~Uszkoreit, L.~Jones, A.~N. Gomez,
  {\L}.~Kaiser, and I.~Polosukhin, ``Attention is all you need,'' in
  \emph{Advances in neural information processing systems}, 2017, pp.
  5998--6008.

\bibitem{miyato2018spectral}
T.~Miyato, T.~Kataoka, M.~Koyama, and Y.~Yoshida, ``Spectral normalization for
  generative adversarial networks,'' \emph{arXiv preprint arXiv:1802.05957},
  2018.

\bibitem{blaauw2019data}
M.~Blaauw, J.~Bonada, and R.~Daido, ``Data efficient voice cloning for neural
  singing synthesis,'' in \emph{ICASSP 2019-2019 IEEE International Conference
  on Acoustics, Speech and Signal Processing (ICASSP)}.\hskip 1em plus 0.5em
  minus 0.4em\relax IEEE, 2019, pp. 6840--6844.

\bibitem{lee2019disentangling}
J.~Lee, H.-S. Choi, J.~Koo, and K.~Lee, ``Disentangling timbre and singing
  style with multi-singer singing synthesis system,'' \emph{arXiv preprint
  arXiv:1910.13069}, 2019.

\bibitem{ping2017deep}
W.~Ping, K.~Peng, A.~Gibiansky, S.~O. Arik, A.~Kannan, S.~Narang, J.~Raiman,
  and J.~Miller, ``Deep voice 3: Scaling text-to-speech with convolutional
  sequence learning,'' \emph{arXiv preprint arXiv:1710.07654}, 2017.

\bibitem{sjolander2003hmm}
K.~Sj{\"o}lander, ``An hmm-based system for automatic segmentation and
  alignment of speech,'' in \emph{Proceedings of Fonetik}, vol. 2003, 2003, pp.
  93--96.

\bibitem{midi}
M.~M. Association, ``Midi manufacturers association,''
  \url{https://www.midi.org}.

\bibitem{toda2007voice}
T.~Toda, A.~W. Black, and K.~Tokuda, ``Voice conversion based on
  maximum-likelihood estimation of spectral parameter trajectory,'' \emph{IEEE
  Transactions on Audio, Speech, and Language Processing}, vol.~15, no.~8, pp.
  2222--2235, 2007.

\end{thebibliography}
